# ItsSQL:
# Intelligent Tutoring System for SQL

*Working Paper*

Sören Aguirre Reid, Frank Kammer, Johannes Kunz, Timon Pellekoorne, Markus Siepermann, Jonas Wölfer

Technische Hochschule Mittelhessen, Wiesenstraße 14, 35390 Gießen, Germany

## Abstract

*SQL is a central component of any database course. Despite the small number of SQL commands, students struggle to practice the concepts. To overcome this challenge, we developed an intelligent tutoring system (ITS) to guide the learning process with a small effort by the lecturer. Other systems often give only basic feedback (correct or incorrect) or require hundreds of instance specific rules defined by a lecturer. In contrast, our system can provide individual feedback based on a semi-automatically/intelligent growing pool of reference solutions, i.e., sensible approaches. Moreover, we introduced the concept of good and bad reference solutions. The system was developed and evaluated in three steps based on Design Science research guidelines. The results of the study demonstrate that providing multiple reference solutions are useful with the support of harmonization to provide individual and real-time feedback and thus improve the learning process for students.*

*Keywords: SQL, Automated Feedback, Intelligent Tutoring System, Individualized Feedback for students, Syntax Distance Measurement*

## Introduction

SQL is the de facto standard for querying relational databases (Wang et al., 2020; Weston et al., 2021; Yang et al., 2021). With the help of SQL, data pools of nearly any size can be handled and analyzed to gain essential insights for decision making (Microsoft, 2023). But although SQL is said to have a relatively small number of commands, mastering SQL is not an easy task. Already small changes in SQL queries can result in an incomplete or wrong solution called *result set* that does not satisfy the initial task. This in turn disturbs the decision-making process and may lead to suboptimal or wrong decisions. Hence, employees with excellent SQL skills are highly sought-after (Weston et al., 2021).

Many different study programs of universities–be it computer science, business science, media systems or else–consider this demand and integrated database courses into their programs. With the increasing demand for STEM graduates, database courses face an increasing number of students (Weston et al., 2021; Wang et al., 2020). This poses some challenges to the teaching process. To be proficient in SQL, practicing is mandatory. Like with any other programming language, students have to work with SQL in databases and solve practical tasks. Concerning the syntax, already the database provides useful hints. But although SQL is much more accessible for students than programming languages (Mitrovic, 1998), students experience many difficulties in the learning process. Particularly when it comes to the correctness of SQL queries in terms of content, students need feedback what they did right and wrong. Providing just a reference solution or the correct result set does not help them to further develop their understanding and skills. Instead, individual feedback is needed. However, with limited staff and an increasing number of students, providing such individual feedback in a one-on-one tutoring situation is hardly feasible (Wang et al., 2018). Manually examining students' submissions (e.g., spotting wrong SQL statements or testing alternative solution approaches) and providing feedback requires a trained eye as well as a careful examination and is very time-consuming (Yang et al., 2021).

To cope with this situation, several intelligent tutoring systems (ITS) have been developed over the past decades which evaluate student submissions and provide feedback to assist students in learning SQL and improving their understanding of SQL concepts. These ITS even have several advantages over traditional methods. They scale to a large number of students (Marin et al., 2017) and provide immediate and personalized feedback to students who do not have to wait for lecturers to assess their submission. However, the effectiveness of these systems highly depends on the accuracy and quality of the feedback provided, which is mostly determined by the approach used for evaluating student submissions (see chapter related literature).

In general, three categories of evaluation approaches can be distinguished (see also next section): static, dynamic, and hybrid approaches (Wang et al., 2020). Static approaches analyze the code of a student submission without executing the code. Usually, static approaches compare student submissions to one or a set of reference solutions and identify the wrong parts of a submission via differences to the reference solutions. The advantage of static approaches consists in the fewer security risks (SQL code can at least delete data from the database) and in the detailed feedback that can be provided. As the code of student submissions is analyzed, faulty parts can be identified and marked accordingly. But the quality of the feedback heavily depends on the number of reference solutions provided and their unambiguity. Often, there exist alternative solutions that are neither defined nor derived by the ITS. Thus, student submissions using these variations are not identified correctly. Anyway, of all ITS using static approaches only one provides detailed feedback to students about their mistakes (e.g., predicates, relations) (Chandra et al., 2016), whereas all other static approaches only provide binary feedback (false/correct) at best (Rivas and Schwartz, 2021 ; Štajduhar and Mauša, 2015).

In contrast to static approaches, dynamic approaches execute the code of a student submission and compare the result set to the result set of the reference solution (Kleerekoper and Schonfield, 2018). If the result sets differ, the submission can be classified as incorrect.

If the results equal, the submission is considered as correct. However, the latter assumption is not necessarily valid. For example, a join of two tables might be avoided if students correctly guess the value of a primary key. Hence, while any correct submission is considered as valid, incorrect submissions might remain unidentified. In addition, dynamic approaches are not able to identify mistakes in student submissions as they solely rely on the comparison of result sets. Therefore, they can only make conclusions based on the submissions' execution behaviour and provide feedback derived from similarities and differences between the result sets (e.g., Wang et al., 2021).

Due to the drawbacks of static and dynamic solutions, more and more hybrid systems have been developed which combine static and dynamic approaches. Interestingly, although hybrid systems promise to overcome the drawbacks of pure static and pure dynamic approaches, most known systems lack of individual feedback (Fabijanic et al., 2020; Wang et al., 2020), cannot analyze incorrect submissions (Chukhray & Havrylenko, 2021) or the authors do not report the type of feedback (Boisvert et al., 2018). However, using a pool of reference solutions has been shown to provide more accurate feedback, as there is also no one-size-fits-all solution for student submissions (Wang et al., 2018; 2020). Even if the two latter systems make use of more than one reference solution, their study does not explain what criteria are used (e.g., distance measurement or harmonization rules) to include a submission as a reference solution and they do not check for the quality of the reference solution.

Thus, ITS for SQL that generate and provide individual feedback to students are still scarce. Even if advanced methods are used, the results of these methods are often not used to give students hints concerning their mistakes. As a result, most systems do not exploit their potential to the full extent. This paper aims at addressing this gap and contributes to the stream of ITS for SQL in the following ways:

First of all, the ITS presented in this paper is a hybrid system that uses static and dynamic evaluation approaches to check the correctness of student submissions and to provide individual feedback that gives detailed hints to students how they can improve their submission. The dynamic evaluation uses a test database with data sets unknown to students so that submissions can be clearly categorized into correct and incorrect solutions in most instances. Incorrect submissions are further analyzed concerning the wrong parts. The insights of these analyses are refined for and provided to students so that they exactly know which parts of their submission are not correct.

Secondly, the static analysis is not only based on one reference solution, but can comprise several solutions. Usually, there exists more than one solution in SQL for a given task. If only one reference solution is used, correct parts of a student submission that does not resemble the reference solution used, but another also correct solution, can be marked as incorrect. This usually confuses students and limits the acceptance of an ITS. Therefore, several reference solutions can be defined, which in turn improves the quality of the feedback. To find the best reference solution suitable to a student submission, a novel distance measurements between two SQL statement is defined.

Thirdly, reference solutions of different quality can be defined. There can be not only more than one solution for a task, but also solutions of different quality (e.g., number of joins, used attributes, length of SQL query etc.). If students submit such solutions, these are accepted as correct, but the students are also told that and how the quality can be improved. For this, not only incorrect submissions are analyzed statically, but also correct solutions, which are also compared against the reference solutions. Without considering solutions of lower quality, these solutions either have to be accepted as correct or rejected as incorrect. In the first case, students do not learn the full potential of SQL. In the second case, students are discouraged because their submission is not really incorrect. Also, they may be confused when their submission is not accepted and they are lead to a different solution which produces the same result set.

Fourthly, our ITS is a self-learning system that continuously updates the set of reference solutions. When analyzing correct student submissions, not only variants of lower quality are found. Usually, alternative solutions of the same quality as the reference solutions exist and are typically found by students. But lecturers are not able to define all solution alternatives. Hence, the ITS identifies such new solutions and integrates them into the set of reference solutions.

Fifthly, the ITS provides lecturers with an easy-to-use user interface where new tasks and the according reference solutions can be defined easily and quickly. A suggestion for a test database is generated automatically so that lecturers are relieved from this time-consuming task.

The remainder of this paper is organized as follows: The next section reviews the existing literature in the field of ITS for SQL and highlights the contribution of this paper in more detail. In the following section, we describe the iterative design of our artifact, based on the design science approach by Gregor and Hevner (2013). Then, we discuss the calculation of suitable distance measurements between two SQL statements and different parts of SQL statements (e.g., SELECT, WHERE, or ORDER BY). To evaluate our artifact, we analyzed a data set of 6914 submissions from 147 students across 27 assignments from the winter term 2021. The paper closes with a discussion of the results, the according implications, and the limitations of this study.

## Related Literature

We have run a search among the data bases AIS, IEEEXplore, and ACM for published journal papers and conference proceedings between 2000 and 2022 with the following search string: ("Structured Query Language" OR "SQL" OR "Relational Database") AND ("automated assessment" OR "analysis" OR "automated grading" OR "learning analytics" OR "feedback"). In total, we identified 103 publications which were screened for the following criteria: (1) the system should analyze SQL queries; (2) the paper should explain the evaluation approach and (3) students should receive (semi- or automated) feedback (e.g., correct or false). In sum, only 15 papers are relevant for our study focusing on SQL. One additional paper who focuses on programming languages uses interesting and reusable methods and is therefore kept in the set of related papers. Seven of these papers use a static approach, two a dynamic, and six use a hybrid approach.

**Static.** One paper (Weston et al., 2021) uses a static clustering approach based on the query structure to cluster student submissions. The approach effectively identifies similar submissions and allows an efficient assessment process. However, static clustering is limited in handling complex syntax or when students use unique solving strategies that may not fit into any cluster. Therefore, the static clustering approach requires either a semi-automated solution alignment or a continuous adjustment of the cluster to add new reference solutions or to refine the statement submitted by the lecturer. With regard to feedback generation, Weston et al. (2021) only provide a dashboard for analyzing student SQL-statement submissions with direct feedback for the lecturer but not for the student. In addition, the clustering approach cannot provide feedback to students concerning syntax.

Two papers employ a static AI-based approach to analyze SQL statements (Rivas and Schwartz, 2021; Štajduhar & Mauša, 2015). In Rivas and Schwartz (2021), an Attention-Based Convolutional Neural Network is trained to classify the correctness of the statement (0/1), provide four remarks (Correct, Partially Correct, Non-Interpretable, and Cheating), and generate a score (0-100). For this purpose, the statement is first tokenized to prepare it for evaluation. Štajduhar & Mauša (2015) use a string similarity method to analyze SQL statements and build a predictive logistic regression model. They compare the entire string of student submissions against a reference solution. Various string similarity methods are utilized to train the model, such as normalized absolute length difference, normalized character Levenshtein distance, normalized word Levenshtein distance, and Euclidean word frequency distance. Based on these distances, the model provides binary feedback, indicating whether the submission is correct or incorrect. Both approaches are limited with regard to the pool of reference solutions. For example, extending the models to new tasks and solution patterns can only be achieved by creating new models. This requires labeling new data and retraining and optimizing the models. Additionally, these models can only handle a limited number of tasks since a large pool of tasks also means a large pool of solution patterns that can overlap due to the many variations in solutions. Therefore, a separate model needs to be created for each task, leading to a significant amount of effort. With regard to the feedback, Štajduhar & Mauša (2015) can only provide binary feedback to students whether it is correct or not correct. Although Rivas and Schwartz (2021) have more categories, they do not provide feedback to students but only to lecturers. Also a specific identification of errors in the submission is missing.

Three papers (Mitrovic 1998 and 2003; Marin et al. 2017) utilize a static rule-based approach. Mitrovic's approach is based on about 500 constraints that describe the fundamental principles of the domain without defining a solution in too much detail, to allow a variety of different solutions (Mitrovic, 1998; 2003).

This aims to teach students declarative knowledge to detect an error, without explicitly naming the mistake (e.g. "Rows were not sorted as expected"). While this approach provides standardized feedback through predefined rules, ensuring students receive consistent feedback, it requires significant domain knowledge and much effort to create the necessary constraints. In contrast, the approach of Marin et al. (2017) offers more flexibility in adding new patterns. This semi-automatic approach requires an instructor to select different patterns for the solution before it can be evaluated using constraints. To add new pattern solutions, either existing constraints need to be recycled and/or new ones must defined. However, Marin et al.'s approach does not provide individual feedback.

Chandra et al. (2016) utilize a tree-based structure approach which converts student submissions into a hierarchical structure. This technique is used on both student submissions and reference solutions. The utilization of tree structures enables the comparison and alignment of discrete segments (e.g., like predicates, relations, group by or having) between the student submissions and the reference solution leading to the identification of errors and their corresponding positions. However, it is not possible to assess the actual executability of the student submissions.

**Dynamic.** Only two papers implement a pure dynamic approach for evaluating SQL statements. In contrast to static approaches, dynamic approaches rely on comparing results obtained from a runtime environment, akin to black-box testing, rather than syntax comparison. This allows an evaluation solely based on the results obtained from executing reference solutions and student submissions. Kleerekoper and Schonfield (2018) use a reference solution to define a task and provide binary automated feedback. In addition, students can see the result sets of their submission and of the reference solution to improve their submission. Wang et al. (2021) compare the result sets of the reference and student solutions with the help of four string similarity metrics (Levenshtein Distance, Jaro-Winkler Distance, FuzzyWuzzy Sorted Token Similarity, and Ratcliff-Obershelp Similarity). Like all dynamic approaches, also these two focus solely on output correctness and do not examine the code structure. Hence, they depend on the test cases used for evaluation and are therefore not well-suited for evaluating the code quality of a submission.

**Hybrid.** Hybrid approaches integrate static and dynamic methods in evaluating student submissions. This promises a more accurate assessment of student work, as it allows for analyzing both query components and their execution behavior in a testing environment. This comprehensive evaluation approach can detect syntactic errors as well as logic and execution-related issues. Boisvert et al. (2018) present a system that allows easy configuration of tasks based on a reference solution and a sample database. During the evaluation, the submission is first statically compared with the keywords from the reference solution. Then both queries (student submission and reference solution) are executed on the sample database, and the results are compared. Based on the comparison, feedback is generated. Unfortunately, the type and quality of the feedback provided is not reported in the paper. Chukhray & Havrylenko (2021) developed an ITS that first statically analyzes the similarity of the student submission to a reference solution. If the similarity is high, they use a syntactic tree comparison to point out mistakes or to verify the submission as correct. If the similarity is low, the query will be executed on a database. Based on the execution results the submission will be marked as correct, or the student has to solve another task from the same topic. In this case the system cannot provide any feedback to the student. Fabijanic et al. (2020) use a combination of constraint-based rule sets with dynamic and static evaluation methods. Based on the constraints the system executes either a dynamic approach e.g., "query can be compiled" or the static tree search approach to e.g., verify if a part of the query is identical to a reference solution. Based on the conditions, feedback is calculated in the form of a grade by subtracting the percentages for each unfulfilled/missed condition. Kleiner et al. (2013) describe a concept in which the student submission is first filtered for forbidden elements (e.g., DROP, DELETE, ALTER) and then checked statically for similarity to the reference solution and code style. In a second step, a dynamic test is processed that calculates the costs in time of the submission and checks the correctness of the result sets with the help of the reference solution and a sample database. Based on this information, individual feedback is generated (e.g., "Datatype of column 2 is wrong" or "Rows were not sorted as expected"). This allows students to receive step-by-step instructions on how to correctly solve a particular task. Wang et al. (2020) propose a comparison of submissions based on previously collected correct student submissions.

For evaluation, they classify the student submissions into three types using a dynamic check: correct statements, non-executable statements, and partially correct statements. Statements that are only partially correct are then transformed into a syntax tree and compared with all other correct statements based on a tree edit distance. Feedback is only provided in terms of grades, but mistakes are not pointed out.

Apart from the aforementioned SQL related approaches, Wang et al. (2018) propose an approach with a pool of reference solutions but for programming tasks. The system determines the most similar reference solution(s) based on syntactic similarity. Reference solutions and student submissions are transformed into abstract syntax trees (ASTs) and then into numerical vectors to calculate the Euclidean distance between each solution and submission. Using a pool of reference solutions increases the likelihood of finding a correct solution, but also has limitations. For example, it is not possible to include every possible correct solution in the pool of reference solutions as the usage of correct but suboptimal solutions can confuse students. Furthermore, there are problems with submissions where the code differs significantly from the reference solution as the approach relies on finding similarities between the student submissions and the reference solutions.

In conclusion, the hybrid approach has been shown to be a more effective and accurate way to assess student work in SQL-related (and also programming) systems than pure systems. By analyzing both the query components and their execution behavior in a testing environment, this approach can detect syntactic errors and logic and execution-related issues, thereby providing a more comprehensive evaluation of student submissions. However, individual feedback is still scarce. Only very few systems consider more than one reference solution. And no system assesses the quality of a submission like we do, i.e., if the code is good or poor.

# Design and Concept of the Artifact

## *Design search process*

For the developed design and concept of the artefact, we choose the Design Science Research Approach described in Gregor and Hevner (2013). This approach includes the following steps: identify problem, define solution objectives, design and development, demonstration, evaluation and communication. There may be several iterations between the different steps, as shown in Figure 1.

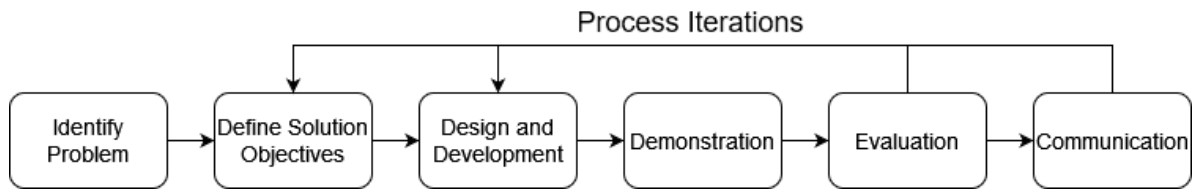

**Figure 1. Design Science Research Approach**

We choose as our evaluation methodology employs an iterative design with three distinct stages, interspersed with test phases. Before delving into the particulars of each stage, we provide a general overview of the process.

The initial stage of our evaluation process, which we refer as "Artifact Step 0", involves dynamic testing to assess the correctness of student submissions. Our implementation uses a fixed test database with a reference solution, to provide binary feedback. This approach focuses solely on the outcome, which has to match the result set of the reference solution. Consequently, this dynamic check serves as our foundational artifact, which was first tested in a database class during the winter term of 2021/22 where the learning process of the students is supported by binary feedback.

In "Artifact Step 1" (hybrid check with one reference solution), we first run the SQL-submission to obtain the right/wrong check. If a solution is incorrect, it is parsed into a static tree. This static tree is then compared to the static tree of the reference solution. This comparison provides indications to each student on where mistakes were made.

In "Artifact Step 2" (multiple reference solutions), multiple reference solutions can be used. We introduce a distance measure between SQL statements to guide the student to the most similar solution.

In "Artifact Step 3" (self-learning), the artifact is extended to automatically learn from correct student submissions and create and update a set of correct solutions semi-automatically so that lecturers do not need to create the pool of reference solutions by themselves which requires significant effort. In addition, submissions and solutions are harmonized so that the set of reference solutions comprises only content-based different solutions.

## *Concept of the Artifact*

**Artifact Step 1.** In the first step we aim to provide better feedback for students on their SQL submissions. On top of the binary evaluation in step 0, we added a static tree comparison like Chandra et al. (2016). The first step in this process is to parse the SQL statement into a syntax tree, from which the keywords and values are extracted. The same process is also performed with the reference solution when the task is created. The keywords as well as the values are then compared. Any discrepancy between the two indicates mistakes made by the student. The student is informed of the mistakes by receiving feedback regarding the incorrect use of attributes, tables and keywords. When using the artifact in practice, it turns out that many different solutions are submitted by students. If only one reference solution is used, the type of feedback in this step forces each student to this reference solution. In other words, mapping a student solution to one reference solution can be a bad idea if there is another correct solution that is much more related to the student submission.

**Artifact Step 2.** We thus extend the artifact in such a way that a lecturer can store several solutions to guide students towards the most similar solution. The reference solutions can be entered directly by the lecturer, or the lecturer can copy a correct student submission into the set of reference solutions. To be able to assign student submissions to a "close" reference solution, a method is needed that evaluates the similarity of two SQL statements. For this reason, we introduce a distance measure between SQL statements to stepwise develop a correct SQL solution. Intuitively speaking, we want our distance measure in such a way that it supports student's learning process best. Hence, the distance measure should meet the following requirement R1:

(R1) The distances of a sequence of student submissions become smaller if the sequence "approaches" a correct solution. In particular, a student submission has a distance of 0 to a reference solution if it is correct. However, vice versa this does not hold.

For the definition of the distance measure respecting R1, we look at a common approach of a student for solving an SQL task and divide them into categories.

*Category C1, used objects.* The first category includes all differences in the use of attributes, tables, variables or values. Differences in this category have the greatest impact on the distance between two SQL statements. This is due to the fact that a student who uses the wrong attributes has fundamental problems in understanding the task. Also, a statement missing an attribute, table or variable usually cannot return a correct result set.

*Category C2, structure.* The second step is to understand the goal of the task and use it to determine the required structure of the SQL statement. Contrary to the differences in category 1, differences in structure can produce the same result sets. It is therefore important to distinguish between different approaches in the structure in order to guide the student to the reference solution that comes closest to the chosen approach. A different approach to solving a task is done by using different SQL constructs. There are also special cases where the use of a different construct should not lead to any distance. (We implemented already some ideas of a harmonization in Step 2, but we did this only for a few special cases and not systematically). For example, the connection of two tables in SQL can be done using a left join. However, it is irrelevant whether two tables T1 and T2 are joined using *"T1 left join T2"* or *"T2 right join T1"*. If the reference solutions contain only one of the joins, the distance measure should be 0 in both cases. The same is the case when using comparison operators. Also here *"a<b"* should have a distance of 0 to the expression *"b>a"*. Not all of these cases can be handled. Here, we limit the distance, because the handling of special cases in the structure makes the system much more error-prone and complex. For example there are "with var"-Constructs which are not considered in our distance measure.

*Category C3, attributes order.* The third category includes the order of used attributes in the select clause as well as in the group by clause.

Mistakes of the different categories have different impacts on the distance. If we have only mistakes from Category 3, the student is already very close to the solution, which should result in a very small distance. Category 1 mistakes instead contain incomplete data and therefore have the largest impact on the distance.

Based on these categories, the distance of a student submission to each reference solution is calculated. The static tree analysis is then executed comparing the syntax tree of the submission and the most similar reference solution (for more details, see Chapter 4). The distance measure thus allows students to be guided to the closest reference solution and receive feedback in direction to that solution.

**Artifact Step 3.** In practice, the artifact of Step 2 suffers from the plethora of different correct solutions that need to be defined by lecturers so that the analysis of student submissions works correctly. Defining all solutions is challenging and a very time-consuming task (see the examples below and the section about harmonization of queries). Additionally, it is likely that lecturers do not discover all different approaches or any tricky solution by a student. Therefore, lecturers need to be assisted in the definition of reference solutions, or better they need to be relieved from this task. Instead, the ITS should learn from the student submissions. However, there are usually many correct submissions that are different, but equivalent. See the example below where either a smaller/greater condition or the between-statement is used to express the same condition.

| Reference solution | SELECT name FROM user WHERE age BETWEEN 18 AND 65; |
|---|---|
| Student submission | SELECT name FROM user WHERE 18 <= age AND age <= 65; |

**Table 1. Example Multiple Correct Submissions**

Standard distance measures (e.g., the Levenshtein distance) cannot be used to identify equivalent solutions. This would inflate the number of reference solutions unnecessarily. What we need is a set of harmonization rules that identify such equivalent statements to give an overview for a lecturer what are the interesting solutions. Therefore, certain rules are applied to the SQL submissions to make similar solutions identical and thus prevent them from being included as a new reference solution. Not all different solutions can be harmonized into a common one. This is the case when we have solutions that use a fundamentally different approach to solving a task. For example, solutions using/not using subqueries should not be harmonized. Another example of such a difference is the use of joins versus the connection of tables by a where clause; see Table 2. Furthermore, it might still be possible to find rules for harmonization, but this can become too time-consuming to resolve complex joins of tables.

| Reference solution | SELECT name FROM user INNER JOIN admin ON user.id = admin.uid; |
|---|---|
| Student submission | SELECT name FROM user, admin WHERE user.id = admin.uid; |

**Table 2. Example Different Approach**

By running our harmonization rules, the distance measure can also be improved. It is now possible to obtain a more accurate distance measure since harmonization means that nearly identical solutions no longer affect the distance, and thus only the actual mistakes or different approaches lead to a higher distance. The improved distance measure results in a new requirement:

(R2) If a correct student submission has a large distance to all reference solutions, it should be added into the set of reference solutions (with high probability).

It cannot always be ruled out that a solution with a high distance measure is also a new reference solution. Therefore, newly added solutions are presented to the lecturer on a dashboard so that s/he can review the admission. By harmonizing it and calculating the distance measure, it is possible to filter out many of the wrong solutions in advance. If this is not possible, we can provide the lecturer with feedback indicating a high distance measure, in particular due to differences in Category C1 (used objects) and Category C2 (structure).

Another improvement that we achieve through applying the harmonization rules is that the number of different solutions is so small that solutions that are actually wrong can be discovered within the small set. Previously, submissions that returned the correct result set were considered correct solutions. However, this allows students to find solutions that work on the test database, but do not return the same result set as the reference solution on every possible database. For example (see Table 3), a student is asked to find the names of all the hotels in the city of York. The student submission returns the same result set as the reference solution, but is actually not correct. If the database did not only contain the hotels of the English cities, but also those of the USA, hotels in the city of New York would also be returned.

| Reference solution | SELECT name FROM hotels WHERE location = "York"; |
|---|---|
| Student submission | SELECT name FROM hotels WHERE location LIKE "%York%"; |

**Table 3. Example Actual Wrong Submission**

If the system detects a wrong solution, it will not be stored in the pool of reference solutions, but saved separately. This way, identical solutions are no longer displayed on the lecturer's Dashboard. If a lecturer marks a solution on the dashboard as incorrect, the system advises him/her to make changes to the database so that this submission no longer leads to a correct solution. He/she can also make and test these changes directly on the dashboard.

# Implementation

To get an overview of the implemented system, the control flow diagram is shown in Figure 1.

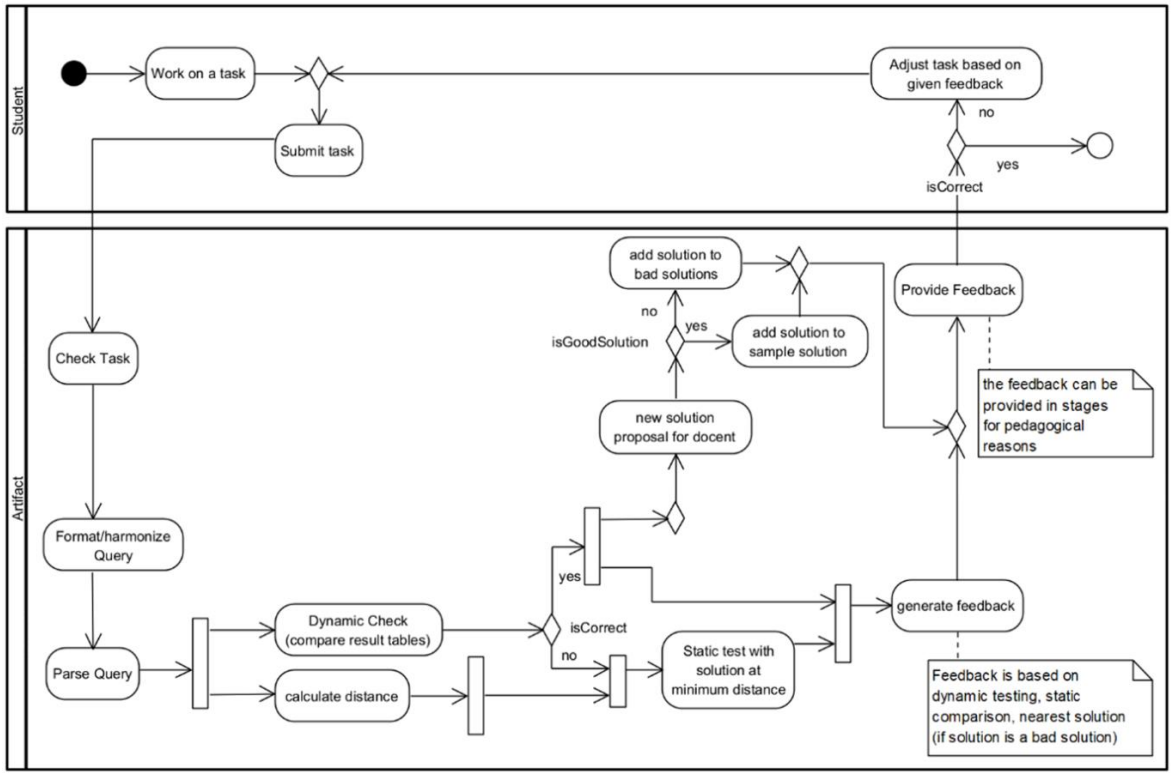

**Figure 1. Control Flow Diagram of the Third Step**

## Static Tree Analysis

For the static tree analysis, the SQL statement is decomposed into the syntax tree. This is then traversed to extract all values and keywords from it. This process is first performed and saved for all reference solutions. If a student now submits an SQL statement, the process is also executed on this. Afterwards, the two trees with their nodes, which contain values and keywords, are compared. The differences are returned to the student as feedback in the form of one or more hints.

## Distance Measurement

To calculate a concrete distance between two statements, we look at the possible differences for the different *clauses* (select, from, where, group by and order by*)* of an SQL statement concerning the previously defined categories.

**Select.** We consider the three categories (C1-C3) to calculate the distance between two *select* clauses. The first category C1 includes the use of the same columns, but in a different order. To define a concrete measure for these differences, the number of required interchanges of attributes is considered, which then form the distance. With regard to structural differences (C2), we only consider the star operator in the *select* clause. Since we can simply harmonize this difference by replacing the star with all the individual columns, the use of a star does not result in any distance to the listing of all columns in detail. For differences of the third category C3, we define the used columns of the statements as a set. Then we calculate the symmetric difference to get the number of the non-matching columns.

**From.** The possible differences in the *from* clause are far more complex than those in the select clause. First, as in the select clause, we also have distances due to the different use of tables. Here the measure is calculated equivalent to the calculation in the *select* clause. In terms of structure, there are two differences to be noted in the *from* clause. The first difference is the use of different joins. The second difference is the use of subqueries as opposed to a simple table reference.

First, we consider the different uses of subqueries. This distance is determined by finding a matching between all subqueries in such a way that the total distance of the matched subqueries is as small as possible. For this we compute a weighted graph where the weights represent the distance of the endpoints. On this graph we then run an algorithm for weighted matching (Edmonds, 1965). If one SQL statement has more subqueries than a second one, then empty SQL statements are used as subquery so that we have the same number of subqueries in both statements.

The calculation of the distance when using different joins is more complex. A harmonization of complex table joins can be very time-consuming. Nevertheless, we want to recognize two different joining operations that have an equivalent semantic and to define such two operations having a distance 0. To achieve this goal, we use the calculation power of database systems. In detail, we use a *test database* with special properties so that only the usage of equivalent operations returns the same result set. The tables of the database consist of only one column. The values in the columns are chosen in such a way that for each subset of tables, we have a value that occurs exactly in that tables. If two SQL statements join tables in an equivalent way, then both statements return the same values.

For example, consider the two statements of Table 4. We have three tables and therefore use numbers from 1 to $2^3-1$. If now both statements are executed on the tables shown in Table 5, both lead to the same result and therefore use an equivalent join of tables.

| Statement 1' | SELECT * FROM T1 LEFT JOIN T2 ON T1.x=T2.x INNER JOIN T3 ON T2.x=T3.x |
|---|---|
| Statement 2' | SELECT * FROM T2 RIGHT JOIN T2 ON T1.x=T2.x INNER JOIN T3 ON T2.x=T3.x |

**Table 4. Example of Two Equivalent Join Statements**

| T1 | T2 | T3 |
|---|---|---|
| x | x | x |
| 1 | 2 | 4 |
| 3 | 3 | 5 |
| 5 | 6 | 6 |
| 7 | 7 | 7 |

**Table 5. Example Test Database**

After creating the tables, an adjusted SQL statement is generated that replaces the tables of the original statement with the tables of our test database and all other parts of the SQL statement are cut off. Thus, the join is executed only on all attributes of the created table and its result set is returned. The details are illustrated in the following example.

| Original query | SELECT sn FROM user LEFT JOIN student ON ... INNER JOIN admin ON ...; |
|---|---|
| Adjusted query | SELECT * FROM a LEFT JOIN b ON a.x=b.x INNER JOIN c ON b.x = c.x; |

**Table 6. Example Adjusted Statement**

To finally calculate the distance measure with respect to the *from* clause of two SQL statements, the size of symmetrical difference of the two result sets of the two adjusted SQL queries is calculated.

**Where.** For calculating the distance measure, the individual conditions must be compared as well as the logical links between them. As the order of the individual conditions can be different, we need to bring them into a uniform structure in order to finally compare them. For this purpose, we translate the conditions into predicate logical formulas, which we then convert into the conjunctive normal form (CNF). This formula can then be lexically sorted and compared with the formula of the other statement. Each different and/or-operator increases the distance by 1.

**Group by/Order by.** The distance with respect to *group by* and *order* clauses is computed similar to the computation of the distance in the *select* clause.

## Query Harmonization

For the execution of the harmonization of the SQL statements described in Chapter 3, we have defined 18 rules for the harmonization based on the data obtained by using Step 1 and StDataep 2 of our artifact. For this purpose, we have analyzed all the correct statements, established rules that allow us to reduce the number of correct, but different SQL statements. Whenever we get a new student submission, each of the rules below is applied to turn the submission into a harmonized SQL statement. However, some harmonization rules reveal complicated or superfluous parts in the SQL statement, e.g., the cast of an integer to integer. In this case, an additional hint is given to the student on how to improve the query. If that statement has a distance of 0 to one of the previous reference solutions, we exclude this as a new reference solution.

---

**Different Aliases**
SELECT c.*    FROM customers c        INNER JOIN orders o        ON c.customer_id = o.customer_id;
SELECT cu.*  FROM customers cu      INNER JOIN orders ord      ON cu.customer_id = ord.customer_id;
**No Aliases  (if possible)**
SELECT c.*             FROM customers c      JOIN orders o       ON c.customer_id = o.customer_id;
SELECT customers.* FROM customers      JOIN orders         ON customers.customer_id = orders.customer_id;
**No Order-by in the Reference-Solution**
SELECT *      FROM customers      ORDER BY customer_id;
SELECT *      FROM customers      ORDER BY customer_id ASC;
**"=" versus "IN" with one-element-set**
SELECT * FROM    employees WHERE salary = (SELECT MIN(salary) FROM employees);
SELECT * FROM    employees WHERE salary IN (SELECT MIN(salary) FROM employees);
**Order-By and LIMIT to realize MIN**
SELECT MIN(salary)    FROM employees;
SELECT salary          FROM employees      ORDER BY salary      LIMIT 1;
**SELECT * instead of SELECT individual attributes**
SELECT *                              FROM employees;
SELECT employee_id, first_name, last_name, [...]    FROM employees;
**Normalize logic expressions in CNF and extract NOT from Predicates**
SELECT *    FROM employees      WHERE        ((salary > 50000 AND division= 'Marketing') OR (salary > 50000 AND division = 'Sales'))    AND employee_Id NOT IN (1, 2, 3);
SELECT *    FROM employees        WHERE        salary > 50000 AND (division= 'Sales'    OR division = 'Marketing') AND NOT employee_Id IN (1, 2, 3);
**Ignore order in projection attributes and ORDER BY if no specific output is required**
SELECT last_name, first_name        FROM employees ORDER BY last_name;
SELECT first_name, last_name        FROM employees;
**Use x <= z AND z <= y instead of z BETWEEN x AND y**
SELECT *    FROM employees    WHERE 50000 <= salary AND salary <= 70000;
SELECT *    FROM employees    WHERE salary BETWEEN 50000 AND 70000;
**Use multiple conditions instead of combined > and < conditions in one comparison**
SELECT *    FROM employees    WHERE salary > 50000 AND salary < 70000;
SELECT *    FROM employees    WHERE 50000<salary<70000;
**">n" and ">=n+1" are replaceable on integer attributes**
SELECT *  FROM employees    WHERE salary > 520;
SELECT *  FROM employees    WHERE salary >= 521;
**Rounding integers to natural numbers is not necessary**
SELECT              COUNT(*)        FROM orders WHERE invoice > 1000;
SELECT ROUND(COUNT(*), 2)      FROM orders WHERE invoice > 1000;
**CAST-Operator not necessary if the data types are already correct**
SELECT *    FROM employees    WHERE age < 30;
SELECT *    FROM employees    WHERE CAST(age as Integer) < 30;
**Use JOIN operator instead of WHERE clause for table relationships**
SELECT *  FROM orders                    JOIN    customers on customer_id = customer_id;
SELECT *  FROM orders, customers    WHERE    customer_id = custome_id:
**Use GROUP BY instead of DISTINCT if no aggregation operations are used**
SELECT DISTINCT name    FROM employees;
SELECT name              FROM employees      GROUP BY name;
**The expressions ISNULL(x) and x IS NULL are equivalent**
SELECT *    FROM employees    WHERE isnull(salary);
SELECT *    FROM employees    WHERE salary IS NULL;

---

| Conditions in WHERE and HAVING can return same query results | | | |
|---|---|---|---|
| SELECT * FROM employees | WHERE division = 'Sales' AND salary > 50000; | | |
| SELECT * FROM employees | HAVING division = 'Sales' AND salary > 50000; | | |
| **Use left and right join replaceable for same join condition** | | | |
| SELECT * FROM employees | LEFT JOIN | divisions | ON employees.div_id = divisions.id; |
| SELECT * FROM divisions | RIGHT JOIN | employees | ON divisions.id = employees.div_id; |

**Table 7. Our Harmonization Rules**

All other solutions are presented to the lecturer in a dashboard (see below) so that the lecturer can quickly decide what are actual wrong solutions and what are new correct solutions. In cases where a solution is likely to be a new solution, it is automatically taken into the set of reference solutions and a lecturer can undo this step with the dashboard. The lecturer also receives information about the solutions to be able to assess its quality more easily. The following dashboard shows an example on the lecturers' information for the SQL task "Calculate the average number of recipe steps". In fact, the lecturer can see five of eight SQL-solutions for that task that remain out of 29 correct submissions entered to our system by the students.

**Figure 2. Lecturer Dashboard for Possible New Solutions.**

### *Applying the concepts to DDL and DML*

The implementation of the artifact on which this paper is based is limited to the evaluation of *select* statements, but many of the concepts can be applied to data definition/manipulation language statements, too.

With regard to the data manipulation language, mistakes can also be divided into the three described categories C1-C3. For example, if we take an *insert* statement (see Table 8), it also has a table name, a set of columns, and a set of values. So the distance of two *insert* statements can be done equivalent to the calculation of the distance measure of the *select* clause by means of the symmetric difference (C3) or the number of interchange operations (C1). The same applies to *update* statements. The difference here is that there is also a *where* clause in these. The calculation of the distance measure or the harmonization of such statements can be done here like the *where* clause of a *select* statement.

| Reference solution | INSERT INTO User (name, age) VALUES ('John Doe', '30'); |
|---|---|
| Student submission | INSERT INTO User (age, surname) VALUES ('30', 'Doe'); |

**Table 8. Insert Statement with two C1 Mistakes.**

Statements of the data definition language differ more in structure. But these can also be divided into different clauses, in which errors can again be assigned to the three categories. In the *create table* clause (see Table 9), for example, a possible error could be an incorrect table name (C3), incorrect (C3) or incorrectly sorted (C1) column names, or a structural difference (C2) due to the use of an existing table and thus a *select* statement.

| Reference solution | CREATE TABLE User (name varchar(255), age int); |
|---|---|
| Student submission | CREATE TABLE User AS SELECT name, age FROM Customer; |

**Table 9. Create Table Statement with a structural difference.**

Even if the implementation does not yet support such statements, it can be stated that an implementation does require an extension of the artifact, but this requires less effort due to the existing concepts.

# Evaluation

## *Usefulness of our Artifact Step 1 and Step 2*

We evaluated an early application of our artifact (Step 1) by analyzing data from the database course in the winter semester 2021/22, in which students received only binary feedback on the entire submission. In this course, students learn basic SQL statements in lectures and must apply their knowledge in 26 SQL tasks. To solve the tasks, students used the artifact and entered a submission for each task until a correct solution was found. This resulted in a dataset of 6914 submissions from 146 students that were used for analysis. In this way, all student submissions to a particular task represent the student's learning process toward the correct solution, with more and more parts of the submission becoming correct.

Now that we have personalized feedback through our hybrid approach, and thus additional information about the submissions, we can analyze the usefulness of our feedback in terms of student learning.

In the following, we always count the number of two consecutive submissions across all tasks and students with respect to specific attributes described below. We also distinguish two cases: In the first case, we count the number of all submissions with respect to only a single reference solution (sr), while in the second case we count the number of submissions with respect to the best matching solution of multiple reference solutions (mr). We found sr=547 and mr=98 cases where student deteriorated a previously correct part of a contribution. That means that when using only one single reference solution, a student's approach is misunderstood in about 1/5 of these cases which shows that when multiple reference solutions are used, the system better understands the student's approach. Instead of guiding the student to a fixed reference solution, the student can be picked up earlier at a crotch of two different reference solutions and are thus better supported on their own learning path to their individual correct solution.

Similarly, we found sr= 13591 and mr= 9484 cases where we did not observe any improvement (sideways movement) between two consecutive submissions. It took an average of sr= 4.01 and mr= 1.61 trials for students to make a submission with progress. Thus, with multiple reference solutions we are able to observe students' progress earlier. We better understand which learning path students want to take. Instead of directing them down a different learning path, we support them in their own approaches to make the learning effect more effective.

## *Usefulness of our Artifact Step 3*

To analyze the harmonization approach, it is compared with one of the standard approaches for measuring the similarity of strings called Levenshtein distance (LD). In particular, the comparison shows how different the various submissions are and how well the harmonization can handle the different SQL queries. On average of all 26 tasks, Figure 2 shows the number of correct submissions that remain if we do not include submissions whose Levenshtein distance is smaller than the x-axes value.

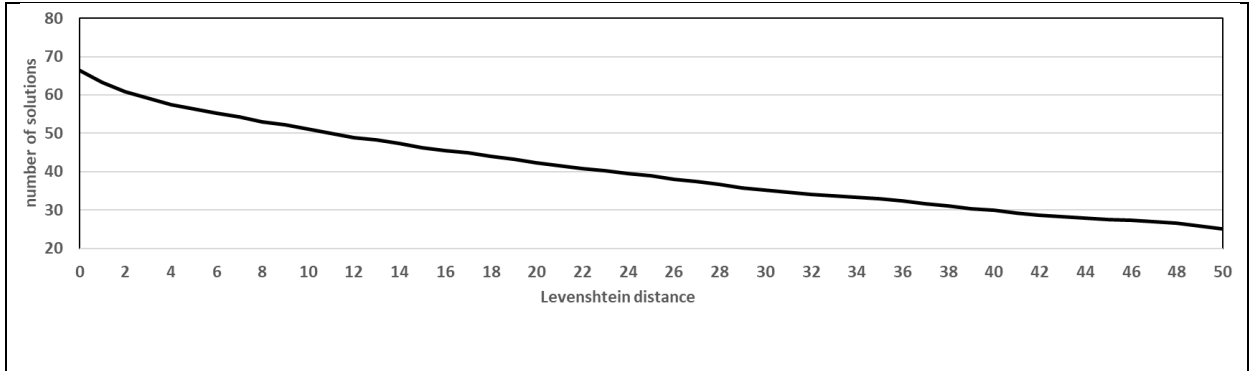

**Figure 3. Solutions Based on the Levenshtein Distance**

There are many different correct solutions, which often follow a similar or the same approach. A pure harmonization based only on the Levenshtein distance can reduce this amount. However, the following example shows that with a Levenshtein distance of 35, we already get two completely different solutions.

| Reference | | Solution |
|---|---|---|
| SELECT  *    FROM cust c      INNER JOIN ord o ON c.cust_id = o.customer_id | | |
| **Levenshtein** | **Distance            =** | **35** |
| SELECT  *    FROM cust c      WHERE cust_id   IN (SELECT o.cust_id FROM ord o) | | |

**Table 10. Example with Levenshtein Distance 35**

We compare the number of remaining solutions after harmonization using our rules with Levenshtein distance of 0 and 35. Over all 27 tasks, our approach of harmonization has on average only 9.0 different solutions compared to 66.44 and 33.0 for Levenshtein distance 0 and 35, respectively. Since these rules take SQL-specific rules into account, we can ensure that only similar solutions are harmonized and that we do not lose any interesting solutions. Specifically, the reduction over the 27 tasks is as follows. We can observe that our harmonization can achieve a significantly higher reduction of solutions in many cases.

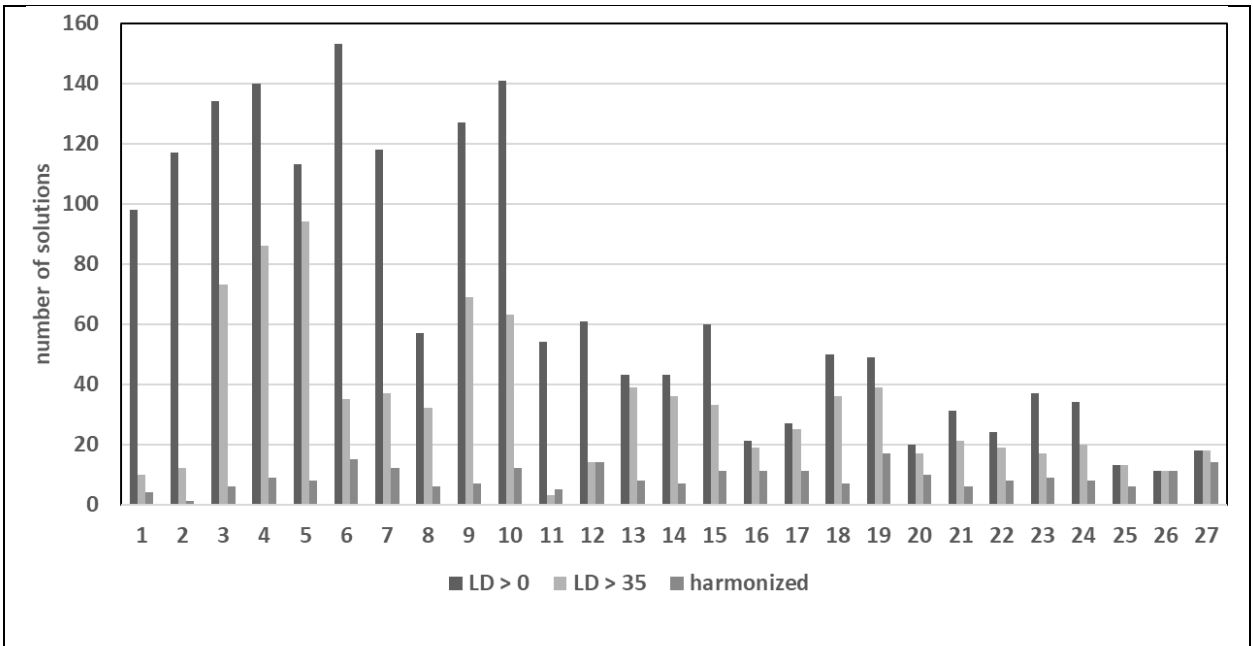

**Figure 4. Solutions Based on the Levenshtein Distance and Harmonization Rules**

# Conclusion

In this study, we presented the development of an artifact that aims at providing personalized feedback to students in a database course. The artifact builds upon an existing system that provides binary feedback from a dynamic test. To improve the effectiveness of the feedback, we implemented improvements in three steps.

In the first step, a static tree comparison was added to provide individual feedback. The goal was to improve the effectiveness of the feedback by identifying syntax differences between the submission and a reference solution. The second step involved the creation of a manually maintained pool of reference solutions. The goal was to provide more diverse and representative reference solutions to support different solution approaches. The third step introduced the harmonization rules to identify equivalent approaches as well as to provide the lecturer a better overview of the different solutions and therefore the ability to detect wrong solutions. The goal was to address the challenge of maintaining a large pool of reference solutions.

The implementation of the first step uses a static tree comparison to perform a syntactical comparison between the student submission and a single reference solution. However, we observed that the feedback generated using this approach was only helpful if the student submission was sufficiently similar to the reference solution. To overcome this limitation, we extended the reference solutions in the second step by creating a manually created pool of reference solutions. We then identified the most similar reference solution to a student submission to provide feedback that was sufficiently comparable to a reference. As a result of our enhancement from Step 1 to Step 2, we have identified that students made fewer backward or sideways moves during their learning process. This is due to the fact that the submissions are now compared to a more similar reference solution, which results in a better ability for the students to adapt the suggestions from the feedback. Providing feedback on the parts that are already correct can also increase students' motivation and boost their confidence in their abilities. However, it is crucial to consider whether the shortest learning path is always the best approach for pedagogical purposes and whether providing extensive feedback to the student immediately is always optimal. People learn from making mistakes.

Hence, giving students the opportunity to make mistakes, to try different approaches without interfering with detailed hints can be more effective concerning the total learning outcome. However, answering this question was not the focus of this study. Nonetheless, having a greater variety of reference solutions is advantageous in improving students' understanding of SQL statements. Rather than imposing a fixed learning path, the ITS should prioritize individualized learning. In this regard, we found that maintaining a large number of reference solutions can be very time-consuming, and it is not always clear when another solution really adds value to the reference pool. To address this issue, we introduced our harmonization rules in a third step to assist lecturers in finding new solutions. In contrast, other systems (e.g., Wang et al. 2018) have encountered difficulties in managing a large pool of reference solutions with numerous variations, resulting in erroneous or untraceable feedback.

## *Implications*

Several lessons can be learned from this study. First of all, we successfully applied the approach of analyzing statements with a syntax tree to the field of SQL statements. As SQL is a programming language, using a syntax tree in ITS for programming languages may be promising. This approach can help to identify those parts of source code that are most probably correct or wrong.

Secondly, in order to provide meaningful support through a comparison with a syntax tree, the use of multiple reference solutions is essential. Students usually follow different solution approaches so that there is no one-size-fits-all reference solution. If an ITS relies only on one reference solution, the creativity of students is curtailed as they are forced to follow only one solution path of many feasible. Whatsoever, finding a suitable reference solution cannot be done with standard measures like the Levenshtein distance but requires a problem-specific distance measure. It is conceivable to transfer also the concept of problem-specific distance measures to the field of programming languages in general. The variety of code for a given task may be even greater than in a descriptive programming language like SQL.

But thirdly, this variety may be reduced by using the concept of harmonization like in this paper. Defining general harmonization rules is most probably easier than defining all possible solution approaches. Often, the amount of different reference solutions is too large to be created manually. A proper set of harmonization rules helps to reduce the number of different solution approaches and to find equivalent solutions. Reducing the number of possible solution approaches is crucial for the maintenance of an ITS because a fully self-learning ITS that continuously updates its set of reference solutions autonomously is still hard to realize. To date, we still need manual intervention of lecturers to finally classify solutions as reference solutions.

So fourthly, making an ITS self-learning is very helpful as it relieves lecturers from defining a plethora of reference solutions. For this, student submissions that have been identified as correct can be employed. However, the quality of correct solutions can widely differ which can be hardly distinguished automatically. Then, the concept of harmonization can help to reduce the amount of different solutions that have to be assessed by lecturers.

Besides programming languages, another application field for problem-specific distance measures and harmonization might be the detection of plagiarism. Harmonization can help to find equivalents while a distance measure can calculate the similarity. As a future research, the likelihood of a plagiarism can be calculated.

### Limitations and Future Research

As always, also this paper and the artifact are not without limitations. First of all, the SQL statements that the ITS can evaluate are limited to statements that produce a result set, i.e., SELECT statements. In the future, other statements of the data manipulation language such as INSERT as well as statements of the data definition language such as ALTER TABLE but also TRIGGER or STORED PROCEDURES are planned to be supported. As described in the Implementation section, some of the concepts developed in this work can be adopted here. Secondly, although the ITS automatically identified new solutions and autonomously updates the set of reference solutions, this process should still be monitored regularly by lecturers. The reason is that although different approaches can lead to the same results, the quality of these approaches can differ widely. In the future, the quality of approaches should also be assessed. This helps either to avoid reference solutions of bad quality, or to grade student submissions.

Thirdly, for the evaluation of the different steps, we evaluated historic practicing data of a database course of the winter term 21/22 where the artifact was at Step 0. Steps 1 to 3 should be analyzed separately. Also, only data from one course was used, which means that comparison groups are missing. In addition, the students' view on the ITS has not been investigated. For future research, the final artifact should be studied for usefulness.

Fourthly, regarding the usefulness, new questions arise concerning the didactics. For instance, how much feedback will be useful/necessary without providing the correct solution and which student needs how much feedback. To this end, the extent to which feedback should be further individualized to best support students' learning should be explored. It should also be considered how far optimal feedback also optimally supports the learning process. It is conceivable that the learning outcome is better when detailed feedback is not given immediately so that students can try several times with intervention from the ITS or a lecturer. Then, students can find solution approaches on their own which improves their knowledge. Future research should focus on these didactical questions.

Fifthly, when using the artifact over several semesters, a lecturer may not want to give the same tasks each semester. For this reason, the system can provide the ability to make simple changes to the tasks to swap attributes, tables or values to provide different tasks for different students.